\documentclass{blois}

\usepackage{color}
\usepackage{colordvi}

\usepackage[normalem]{ulem}

\usepackage{color}

\newcommand{\beqn}{\begin{eqnarray}}
\newcommand{\eeqn}{\end{eqnarray}}
\newcommand{\be}{\begin{equation}}
\newcommand{\ee}{\end{equation}}

\newcommand{\ba}{\begin{array}{c}}
\newcommand{\bat}{\begin{array}{cc}}
\newcommand{\ea}{\end{array}}

\newcommand{\bi}{\begin{itemize}}
\newcommand{\ei}{\end{itemize}}

\newcommand{\ket}{\,\rangle}
\newcommand{\bra}{\langle \,}

\newcommand{\Frac}[2]{\frac{\displaystyle #1}{\displaystyle #2}}

\newcommand{\cO}{{\cal O}}

\newcommand{\mL}{\mathcal{L}}
\newcommand{\mM}{\mathcal{M}}
\newcommand{\mO}{\mathcal{O}}

\newcommand{\Int}{\displaystyle{\int}}

\newcommand{\bear}{\begin{eqnarray}}
\newcommand{\eear}{\end{eqnarray}}
\newcommand{\nn}{\nonumber}

\newcommand{\Photo}{\includegraphics[height=35mm]{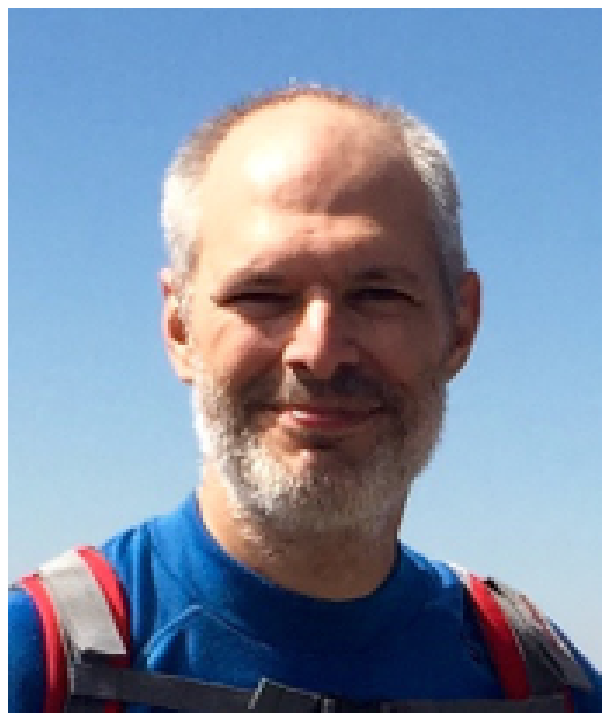}}

\begin{document}

\begin{flushright}
IFT-UAM/CSIC-15-102   \\
FTUAM-15-29
\end{flushright}
%
\vspace*{1.5cm}
\title{Electroweak chiral Lagrangian with a light Higgs}

\author{ Juan Jos\'e Sanz-Cillero~\footnote{
Talk given at 27th Rencontres de Blois, May 31 - June 5 2015, Blois, France.
I would like to thank C.~Grojean and the organizers of Rencontres de Blois for their invitation and the kind scientific
environment enjoyed at Blois.
This work has been supported by ERDF funds from the European Commission
[FPA2013-44773-P, SEV-2012-0249, CSD2007-00042].    
} }

\address{
Departamento de F\'\i sica Te\'orica and Instituto de F\'\i sica Te\'orica, IFT-UAM/CSIC,
Universidad Aut\'onoma de Madrid, Cantoblanco, 28049 Madrid, Spain }

\maketitle\abstract{
In this talk we discuss the structure of electroweak low-energy effective theories
where the Higgs is non-linearly realized, typically in scenarios where the Higgs is a pseudo Nambu-Goldstone boson (pNGB)
of some beyond Standard Model (BSM) symmetry.
The organization of the perturbative counting and the relevance of the various
next-to-leading order contributions is studied. We discuss some new results on the structure
of the one-loop ultraviolet divergences and the contribution from tree-level heavy resonance exchanges to the low-energy
effective theory, which are applied to a couple of explicit examples in order to show how,
in the non-linear effective theory --the electroweak chiral Lagrangian with a light Higgs (ECLh)--, one-loop corrections
can be as important as the contribution from higher dimension operators.
}

\section{Chiral power counting in non-linear effective theories}

There are two usual approaches to
low-energies electroweak (EW) effective field theories (EFT) according to how
the Higgs $h$ and the EW Goldstones $\omega^a$ are introduced:
\begin{enumerate}

\item
{\bf Linear EFT:} The Higgs $h$ and EW Goldstone $\omega^a$ fields
conform a complex doublet $\Phi$ of the EW symmetry.

\item
{\bf Non-linear EFT:} The Higgs field $h$ is introduced as a singlet and the EW Goldstones are
non-linearly realized through the unitary matrix $U(\omega^a)$.

\end{enumerate}
The non-linear approach is indeed more general: it includes also the linear-Higgs EFT, as one can always write down
the doublet $\Phi$ in its polar form in terms of the modulus $(v+h)/\sqrt{2}$ and a unitary matrix $U(\omega^a)$.
The non-linearity of the model is indeed a quality that is related to the separation from the linear scenario rather
than whether one chooses to express $h$ and $\omega^a$ in a non-linear way.
Many BSM frameworks show this non-linear structure,
e.g., models where $h$ is a pNGB~\cite{composite-rev}
and, hence, it transforms non-linearly under the spontaneously broken generators of the BSM symmetry.


The non-linear EFT Lagrangian~\cite{Longhitano:1980iz,ECLh-Gavela,EW-chiral-counting}
is sorted out according to the ``chiral'' dimension of its
operators~\cite{EW-chiral-counting,Weinberg:1979+Georgi-Manohar,Hirn:2004,1loop-AA-scat,EW-resonance},
not through their canonical dimension which one uses in linear EFT's:
\bear
\mL_{ECLh} &=& \mL_{p^2} \,+\, \mL_{p^4}\, +\, ...
\eear
where derivatives and masses of the particles count as $\cO(p)$, $h/v$ count as $\cO(p^0)$
--and the same occurs with other boson fields--,
and fermion fields scale like $\psi/v\sim \cO(p^{1/2})$.
$v=(\sqrt{2} G_F)^{-1/2}=246$~GeV is the Higgs vacuum expectation value.
A simple dimensional
analysis~\cite{EW-chiral-counting,Weinberg:1979+Georgi-Manohar,Hirn:2004,1loop-AA-scat,EW-resonance}
shows that the amplitudes have an expansion of the form (e.g. for a $2\to 2$ process)
\bear
\mM &\sim & \underbrace{ \Frac{p^2}{v^2}  }_{\mbox{LO (tree)}}
\, + \, \bigg(
\, \underbrace{ c_{k}^r }_{\mbox{ NLO (tree) }} \quad -\quad
\underbrace{   \Frac{ \Gamma_{k,n} }{16\pi^2}\ln\Frac{p}{\mu} \quad +\quad ...  }_{
\mbox{NLO (1-loop)}    }
\quad
\bigg)
\,\,\, \Frac{p^4}{v^4}\, \,\, +\,\,\, \cO(p^6) \, ,
\eear
The three singled-out contributions have different origins:
\begin{itemize}

\item
{\bf The LO amplitude} is given by the tree-level diagrams
provided by the vertices from the LO Lagrangian $\mL_{p^2}$.

\item
{\bf The NLO amplitude} has two types of contributions:

\begin{enumerate}
\item
{\bf One-loop diagrams} with vertices only from the LO Lagrangian $\mL_{p^2}$.
These contributions are typically suppressed with respect to (wrt) the LO in the form
$p^2/\Lambda_{\rm non-lin}^2$,
with this scale
$\Lambda_{\rm non-lin}\sim 4\pi v$
directly related to the non-linearity of the BSM scenario:
these corrections vanish when the Higgs can be
linearly realized through a complex doublet $\Phi$~\cite{Guo:2015}.

\item
{\bf Tree-level diagrams}  with one vertex of higher dimension, from $\mL_{p^4}$.
In the underlying BSM theory these constants can get contributions from tree-level exchanges of heavy states
of mass $M_R$ not included in the EFT~\cite{EW-resonance}.
Although the renormalized EFT couplings also get corrections from resonance loop diagrams,
the suppression of these NLO corrections can be estimated as $c_k^r p^2/v^2 \sim p^2/M_R^2$.

\end{enumerate}
\end{itemize}
In the case when the scale $\Lambda_{\rm non-lin}$ that governs the non-linearity
is much higher than the masses $M_R$
of the intermediate heavy states
NLO loops are highly suppressed wrt NLO tree-level corrections
and the linear-Higgs EFT approach is more appropriate.
However, in the case when both scales $\Lambda_{\rm non-lin}$ and $M_R$ are similar,~\footnote{
In this cases the $\cO(p^4)$ couplings are of the order of
$c_k^r\sim (16\pi^2)^{-1}$~\cite{Weinberg:1979+Georgi-Manohar,chpt}.
}
one must account for both the tree-level and one-loop NLO corrections
to have reliable determinations of the observables at that precision
and the linear-Higgs EFT is then inappropriate.

\section{One-loop NLO corrections: ultraviolet divergences}

The background field method in path integral allows one to study the loop corrections
to the effective action.
Expanding the Lagrangian $\mL_{p^2}$ in powers of the fluctuation $\vec{\eta}^T =(\Delta^a,\epsilon)$
(with $\Delta^a$ and $\epsilon$ providing the Goldstone and Higgs fluctuations, respectively)
around the solutions of the equations of motion (EoM)
one has obtains~\cite{Guo:2015}:
\bear
\mL_{p^2} &=&
\underbrace{ \mL_{p^2}^{\cO(\eta^0)} }_{ \mbox{tree-level}}
\,\,\,+\,\,\,
\underbrace{ \mL_{p^2}^{\cO(\eta^1)} }_{ \mbox{EoM}}
\,\,\,+\,\,\,
\underbrace{ \mL_{p^2}^{\cO(\eta^2)} }_{ \mbox{1-loop}}
\,\,\,+\,\,\,
\underbrace{  \cO(\eta^3)  }_{ \mbox{higher loops}} \, ,
\eear
where the $\cO(\eta^0)$ term yields the tree-level diagrams with LO vertices,
the requirement that the linear term vanishes provides the EoM at LO
and $\mL_{p^2}^{\cO(\eta^2)}$ gives the one-loop NLO amplitude~\cite{Guo:2015,book-Donoghue}.
The remaining terms provide the amplitudes
at two loops and higher, i.e., at next-to-next-to-leading order (NNLO) and higher
in the chiral expansion~\cite{book-Donoghue}.
$\mL_{p^2}^{\cO(\eta^2)}$ can be rearranged as the quadratic form
\bear
\mL_{p^2}^{\cO(\eta^2)} &=& - \Frac{1}{2} \vec{\eta}^T\, (d_\mu d^\mu + {\rm\bf  \Lambda})\vec{\eta}\, ,
\eear
in terms of the corresponding operators ${\rm\bf \Lambda}$ and $d_\mu$ operators
determined by the structure of $\mL_{p^2}$~\cite{Guo:2015}.~\footnote{
It is instructive to observe that from the chiral counting point of view ${\rm\bf \Lambda}\sim\cO(p^2)$
and $d_\mu \sim \cO(p)$~\cite{Guo:2015}.
}
The integration of this term over the fluctuations $\vec{\eta}$
yields
\bear
S^{1\ell} &=& \Frac{i}{2}    {\rm tr}      \, \log{ \left(d_\mu d^\mu +  {\rm\bf \Lambda}  \right)}
\,\,\,=\,\,\, -\Frac{\mu^{d-4}}{16\pi^2 (d-4) }\Int {\rm d^d x} \bra \Frac{1}{12} [d_\mu ,d_\nu] [d^\mu ,d^\nu]
+\Frac{1}{2}{\rm\bf \Lambda}^2 \ket + {\rm finite}
\nn\\
&&\hspace*{3cm} = -\Frac{\mu^{d-4}}{16\pi^2 (d-4) }\Int {\rm d^d x} \sum_k \Gamma_k \mO_k + {\rm finite} \, ,
\eear
where $\bra ..\ket$ stands for the matrix trace,
and $\mu$ is the renormalization scale and $d$ is
the space-time dimension in dimensional regularization. The 1st line has been reexpressed in the 2nd line in
terms of the basis of EFT operators $\cO_k$.
The ultraviolet (UV) divergences are cancelled out by means of appropriate $\cO(p^4)$ counter-terms of the form
\bear
\mL_{p^4}^{\rm ct} &=& \sum_k c_k \mO_k\, , \qquad\mbox{with the renormalizations } \,
c_k \,= \, c_k^r(\mu) \,+\, \Frac{\mu^{d-4}\, \Gamma_k}{16\pi^2 (d-4)}\, .
\eear
The couplings of the EFT Lagrangian that get renormalized due to Higgs and EW Goldstone loops
are listed in Ref.~\cite{Guo:2015}.
%

Of course, though important, this is not the end of the story in non-linear EFT's:
one must compute the full one-loop amplitude for every particular process under study,
with the full structure of logs, polylogs and rational~\cite{loop-calcs}
pieces, not just the running~\cite{Guo:2015}.

\section{Tree-level NLO corrections: predictions from composite resonance exchanges}

The exchange
of heavy resonances leads at low energies to NLO and higher order EFT operators suppressed
by powers of $p/M_R$~\cite{EW-resonance,rcht}.
At NLO one only needs the resonance Lagrangian compatible with the SM symmetries
that are linear in the resonance fields $R$ and have at most two derivatives
(or analogous light scales $p$)~\cite{EW-resonance,rcht}.
For instance, in the case of a parity preserving strongly coupled model,
the resonance Lagrangian with triplet vector ($V$) and axial-vector ($A$) multiplets
that is relevant for $\mL_{p^4}$ has the form~\cite{EW-resonance}:
\bear
\mL_V+\mL_A = \sum_{R=V,A}\bra R_{\mu\nu} \chi_R^{\mu\nu}\ket\, ,
\eear
with $\chi_V^{\mu\nu} = \Frac{F_V}{2\sqrt{2}}f_+^{\mu\nu}
+\Frac{i G_V}{2\sqrt{2}}[u^\mu,u^\nu]\, +\, ...$
and
$\chi_A^{\mu\nu} = \Frac{F_A}{2\sqrt{2}}f_-^{\mu\nu}
\, +\, ...$
The spin--1 resonances $R_{\mu\nu}$ are described in the antisymmetric tensor formalism~\cite{rcht},
and one has the tensors $f_\pm^{\mu\nu}= u^\dagger \hat{W}^{\mu\nu} u \pm u \hat{B}^{\mu\nu} u^\dagger$
given by the field strengths of the $W_\mu^a$ and $B_\mu$ fields
and $u_\mu=i u D_\mu U^\dagger u$, with $u^2 =U(\omega^a)$~\cite{Pich:2013}.
$\chi_R^{\mu\nu}$ contains only particles in the low-energy EFT.
Integrating out the spin--1 resonances at low energies, one obtains
the tree-level contribution to the $\cO(p^4)$ ECLh~\cite{EW-resonance}
\bear
\mL_{p^4}^{\rm from \, V,A} &=& -
\sum_{R=V,A} \Frac{1}{M_R^2}
\left( \bra \chi_R^{\mu\nu} \chi_{R\, \mu\nu}\ket - \Frac{1}{2}\bra \chi_R^{\mu\nu}\ket^2\right)
=
- \Frac{i F_V G_V}{4 M_V^2}\bra f_+^{\mu\nu} [u_\mu, u_\nu ]\ket + ...
\eear
obtaining a prediction
for the $\cO(p^4)$ constants in terms of the $V$ and $A$ masses and couplings~\cite{EW-resonance}
\bear
a_1\big|_{V,A} =& - \Frac{F_V^2}{4 M_V^2} + \Frac{F_A^2}{4 M_A^2} &\, \,\,\stackrel{\rm UV\, compl.}{=}
\,\,\, -\, \Frac{ v^2}{4} \, \left(\Frac{1}{M_V^{2}} + \Frac{1}{M_A^{2}}\right)\, ,
\nn\\
a_2-a_3 \big|_{V,A} =& \, -\, \Frac{F_V G_V}{2 M_V^2}  & \, \,\,
\stackrel{\rm UV \, compl.}{=} \, \,\, -\,\Frac{v^2}{2 M_V^2}\, ,
\eear
where we have used some UV-completion hypotheses in the last equalities of each line:
in the case of $a_1$ we assume the strongly coupled theory is asymptotically free and one has the
two Weinberg sum-rules $F_V^2=F_A^2+v^2$ and $F_V^2 M_V^2=F_A^2 M_A^2$~\cite{WSR,Peskin:92};
in the case of $a_2-a_3$ the electromagnetic
form-factor into two composite EW Goldstone bosons in the theory with resonances
is ${\rm\bf F}_{\gamma^*\omega\omega}(s)=1 + (F_VG_V/v^2) \, s/(M_V^2-s)$
and demanding that it vanishes at infinite momentum transfer $s\to \infty$ yields the constraint
$F_VG_V=v^2$ used above~\cite{EW-resonance}.
%

\section{The importance of being earnest and keeping the full NLO: two examples}

The importance of the different NLO pieces becomes obvious if one consider the previous
asymptotically-free strongly coupled benchmark scenario.
At one-loop in the resonance theory the oblique parameters
lead to the constraint $M_A^2=M_V^2/\kappa_W$
and the 95\% confidence level determinations $0.94<\kappa_W<1$, $M_V>4$~TeV~\cite{Pich:2013}.~\footnote{
The $h\to WW$ coupling is normalized such that $\kappa_W^{\rm SM}=1$.
}
Taking the lower allowed limits for $M_V$ and $\kappa_W$  (which produce
the maximal possible effects)
one gets the predictions
\begin{itemize}

\item
{\bf Oblique S--parameter (NLO tree dominance):}
at NLO one has the structure~\cite{1loop-AA-scat}
\bear
S &=& - 16 \pi \, \left[
\underbrace{a_1^r(M_V)}_{{\rm Tree\,} \approx - 2\times 10^{-3}} +
\underbrace{ \Frac{\kappa_W^2-1}{192\pi^2} }_{ {\rm loop\, } \approx - 6\times 10^{-5} }
\left(\ln\Frac{M_V^2}{m_h^2}+\Frac{5}{6}\right)\right] \, ,
\eear
with $a_1^r(\mu) = - v^2/4 (M_V^{-2} + M_A^{-2}) +
\{ [8/3+\ln(\mu^2/M_V^2)] - \kappa_W^2 [8/3 +\ln(\mu^2/M_A^2) \}/(192\pi^2)$.~\cite{Pich:2013}


\item
{\bf $\gamma\gamma\to W^+W^-$ scattering (NLO loop dominance):}
in the energy region
$m_{W,Z,h}^2\ll s \ll \Lambda_{ECLh}^2$
this amplitude is given at NLO by the scalar function~\cite{1loop-AA-scat}
\bear
A_{\rm NLO}^{\gamma\gamma\to \omega\omega} &=& \Frac{1}{v^2}
\left[  2 \kappa_W c_\gamma^r  + \underbrace{ 8(a_1^r -a_2^r + a_3^r) }_{
{\rm tree\, }\approx 0.5\times 10^{-3} }
+  \underbrace{ \Frac{\kappa_W^2-1}{8\pi^2}  }_{ {\rm loop\, }\approx -1.5\times 10^{-3}  }
\right]
\, ,
\eear
with the $a_1^r-a_2^r+a_3^r$ estimate from the previous Section.
The value of the $h\to \gamma\gamma$ $\cO(p^4)$ coupling $c_\gamma^r$
is in principle undetermined in our analysis~\cite{EW-resonance}. If the impact of $c_\gamma^r$ is
as important as that from $a_1^r-a_2^r+a_3^r$ one realizes that the one-loop correction
is more important that the tree-level NLO amplitude
and should not be dropped.

\end{itemize}

\end{document}